Data Resource Profile:  Egress Behavior from Select NYC COVID-19 Exposed Health Facilities March-May 2020

*Debra F. Laefer, Center for Urban Science and Progress and Department of Civil and Urban Engineering, New York University, 370 Jay St. #1303C, Brooklyn, NY 11201, USA, debra.laefer@nyu.edu

Thomas Kirchner, School of Global Public Health, New York University

Haoran (Frank) Jiang, Center for Data Science, New York University

Darlene Cheong, Department of Biology, New York University

Yunqi (Veronica) Jiang, Tandon School of Engineering, New York University

Aseah Khan, Department of Biology, New York University

Weiyi Qiu, Courant Institute of Mathematical Sciences, New York University

Nikki Tai, Courant Institute of Mathematical Sciences, New York University

Tiffany Truong, College of Arts and Science, New York University

Maimunah Virk, College of Arts and Science, New York University

Word count: 2845

**Abstract**

**Background:** Vector control strategies are central to the mitigation and containment of COVID-19 and have come in the form of municipal ordinances that restrict the operational status of public and private spaces and associated services. Yet, little is known about specific population responses in terms of risk behaviors.

**Methods:** To help understand the impact of those vector control variable strategies, a multi-week, multi-site observational study was undertaken outside of 19 New York City medical facilities during the peak of the city's initial COVID-19 wave (03/22/20-05/19/20). The aim was to capture perishable data of the touch, destination choice, and PPE usage behavior of individuals egressing hospitals and urgent care centers. A major goal was to establish an empirical basis for future research on the way people interact with three-dimensional vector environments. Anonymized data were collected via smart phones.

**Results:** Each data record includes the time, date, and location of an individual leaving a healthcare facility, their routing, interactions with the built environment, other individuals, and themselves. Most records also note their PPE usage, destination, intermediary stops, and transportation choices. The records were linked with 61 socio-economic factors by facility zip code and 7 contemporaneous weather factors and then merged in a unified shapefile in an ARCGIS system.

**Conclusions:** This paper describes the project team and protocols used to produce over 5,100 publicly accessible observational records and an affiliated codebook that can be used to study linkages between individual behaviors and on-the-ground conditions.

**Key Features**

- This publicly accessible data set of 5,163 records provides the detailed movement, touch, and PPE usage of individuals leaving New York City (NYC) healthcare facilities in Spring 2020.
- Trip duration and records of transportation and destination choices are noted.
- This 9-week study recorded perishable data outside of 10 hospitals and 9 urgent care clinics in 4 of NYC's 5 boroughs.
- The described protocols provide an empirical basis for future research on the way people interact with three-dimensional vector environments.

**Data Resource Basics**

The spread of infectious disease through a population is inherently a spatial and temporal process. Classic approaches to transmission modeling focus on the clustering of people in space and time, wherein person-to-person contact produces new infections among susceptible hosts. Yet classic susceptible–infected–recovered (SIR)[1] and derivative epidemic modeling approaches do not attempt to distinguish between the physical characteristics of the places people congregate nor the specific ways that people interact with these places or with others in the scene. Mitigation of COVID-19 requires minimizing surface-vector transmission and person-to-person contact, not just limiting the total number of people congregating. Vector control strategies in public health involve efforts to contain or mitigate the spread of disease by intervening upon the "vectors" — i.e., modes of transportation — that carry disease "agents" to their destination (susceptible "hosts"). When disease transmission goes viral, as with COVID-19, members of the population serve as both susceptible hosts and as vectors for transmission of the disease agent to others. Over the past nine months various governmental efforts to implement vector control measures have involved limiting opportunities for people to engage with physical spaces and social environments, including an array of "surface vectors" like door handles, glassware, personal protective equipment (PPE), and other objects with which people physically interact. The specific points of success and failure within these policies have not been systematically documented. As such, effective further intervention and targeting messaging are hard to devise. Understanding the effectiveness of control measures is critical to managing the current wave of this epidemic. Thus, a data collection effort funded by the National Science Foundation (NSF) was undertaken to collect "perishable" data around the initial implementation of the governmental PAUSE[2] ordinance during New York State's spring COVID-19 peak. The observational study was intended to establish a protocol and a short, longitudinal study of various behaviors of individuals leaving medical facilities during high COVID-19 infections rates. Specifically, note was made of destinations, transportation choices, touch behaviors and PPE usage, as well as the individual's gender. The nine-week study covered 19 facilities across four of New York City's five boroughs (Queens, Brooklyn, Manhattan, and the Bronx). Additional funding is being sought to extend the scope and duration of this study. The study was deemed as exempt under New York University's institutional review board (IRB-FY2020-4305).

**Data Collected**

The data collection protocols were designed to collect anonymized, hyper-local behavioral choices that are not possible to ascertain from cell phone, footfall data, closed-circuit television, or traffic cam data. To achieve this, 16 New York University students were recruited as observers immediately prior to New York's implementation of the PAUSE order on the evening of Mar. 22, 2020. The study was funded for 9 weeks with student observers collecting data 10-20 hours per week (class schedule permitting) until May 19, 2020. Data were collected via smartphones.

The intent was to have an even distribution of observers between hospitals and urgent care clinics in multiple New York City boroughs. Because of the critical time constraints on the project, student observers were interviewed and then hired on a first come, first serve basis with no more than 1 student hired per zip code. Ultimately, facilities were selected based on the ability of the observer to reach the selected location by foot.

Each observer was assigned one location. Soon after, adjustments were made because of low visitor rates in two instances (one urgent care in Brooklyn and one in Queens) and poor viewing logistics at two others (one hospital in the Bronx and one in Manhattan). This resulted in four of the 19 sites having very few records, as the observers were reassigned. Ultimately, records were collected at 17 facilities across. In Manhattan, two urgent care facilities were located directly across the street from each other on West 42nd Street, they are reported as a joint reporting location. All facilities are shown in Figure 1 superimposed atop COVID-19 infection rates per zip code[3]. These hospitals and urgent care clinics were situated in both highly infected neighborhoods and those showing less than 50% of the city's average infection rate.

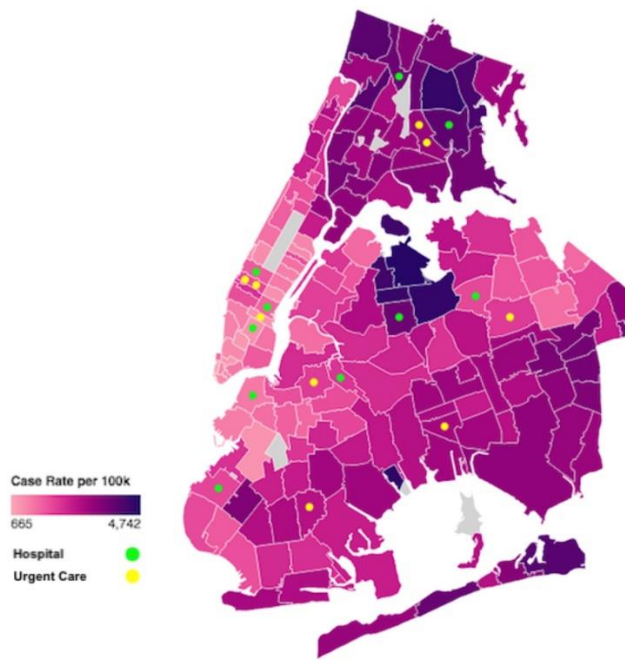

Figure 1. Schematic representation of the locations of the 19 medical facilities superimposed on COVID-19 infection rates per zip code based on infection rate data as of July 29, 2020

Generally, there was one observer per facility. However, in the case of the NYU Langone hospital in Brooklyn, there were ultimately two observers (situated at different entrances) due to the closure of a previously monitored urgent care clinic. Over 1,500 hours of data were collected at all times of the day and all days of the week in the period of March 22-May 19, 2020. Ultimately this resulted in the collection of 5,163 records. However, due to the ad hoc nature of the data collection, which depended upon facility hours and student availability, there is unevenness in the quantity of records (Fig. 2) and their temporal distribution (Fig. 3).

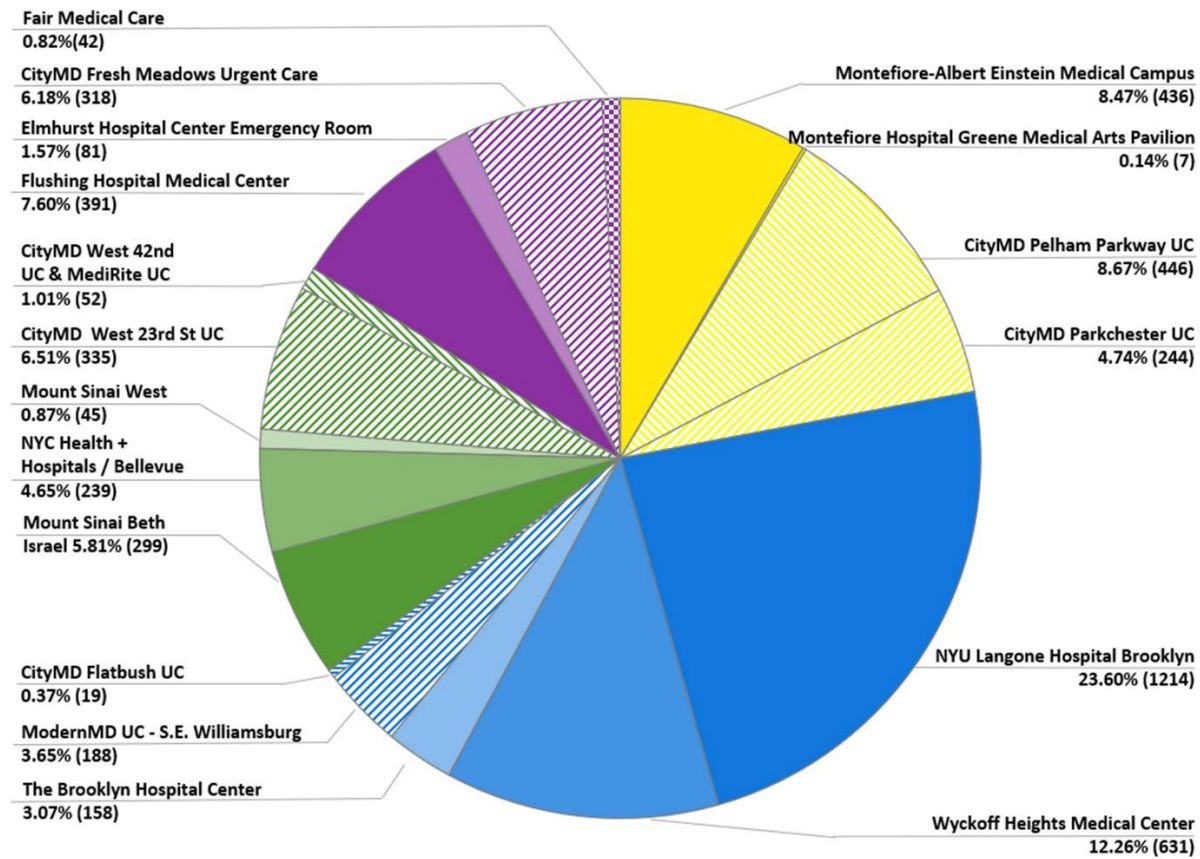

Figure 2. Cumulative distribution of available records

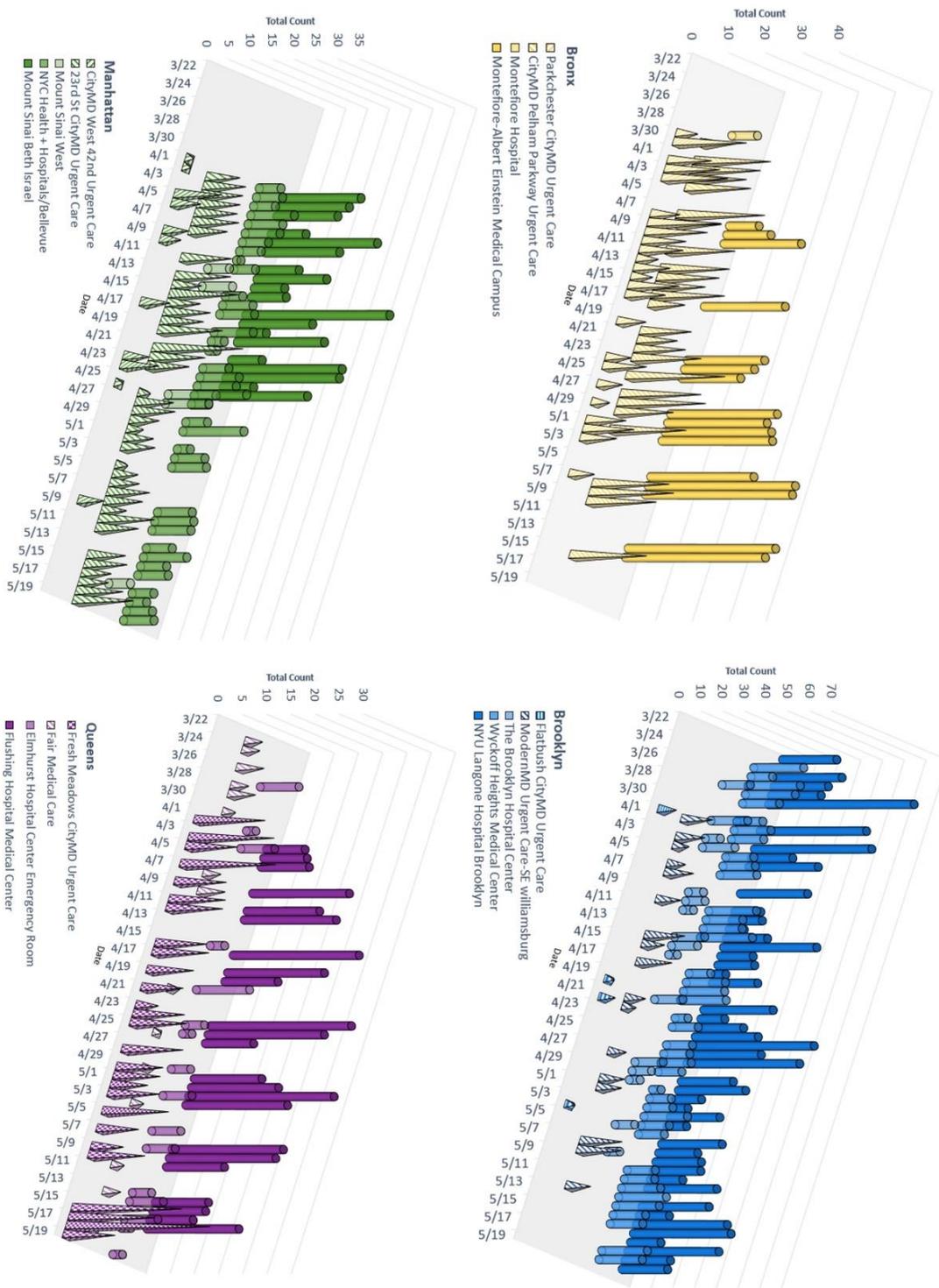

Figure 3. Temporal distribution of the data by collection date in 4 of NYC's boroughs

The observers were stationed across the street from their facilities to keep out of harm's way and instructed to randomly select a subject, as a person exited the designated medical facility. No consideration was given or notation made as to whether the individual was likely to be medical staff, patient, or visitor. The subject was followed from across the street until one of three outcomes occurred: (1) the subject entered a vehicle, subway station, or building and was no longer visible; (2) the subject walked more than 1.3 km from the medical facility; or (3) tracking exceeded 20 minutes.

On smartphones, the observers traced the subject's route and noted locations of interactions with the built environment or other individuals. The route duration and time of day were recorded, as well as the subject's gender and PPE usage. The record also noted the position and nature of objects touched, locations visited, and if applicable, transportation choice (e.g. bus, subway, taxi, personal vehicle, by foot). Locations visited were noted by facility type (e.g. coffee shop, pharmacy, deli, food truck), and special note was made of individuals returning to the medical facility or entering a nearby one (e.g. temporary tent, adjacent building). On iPhones, the data were collected via DrawMaps. On Androids the MyMaps app was used. The observations were recorded without photographs, video, or oral interaction with the subjects to preserve anonymity. Sample records are shown in Fig. 4. The typical record lasted 5 minutes in duration.

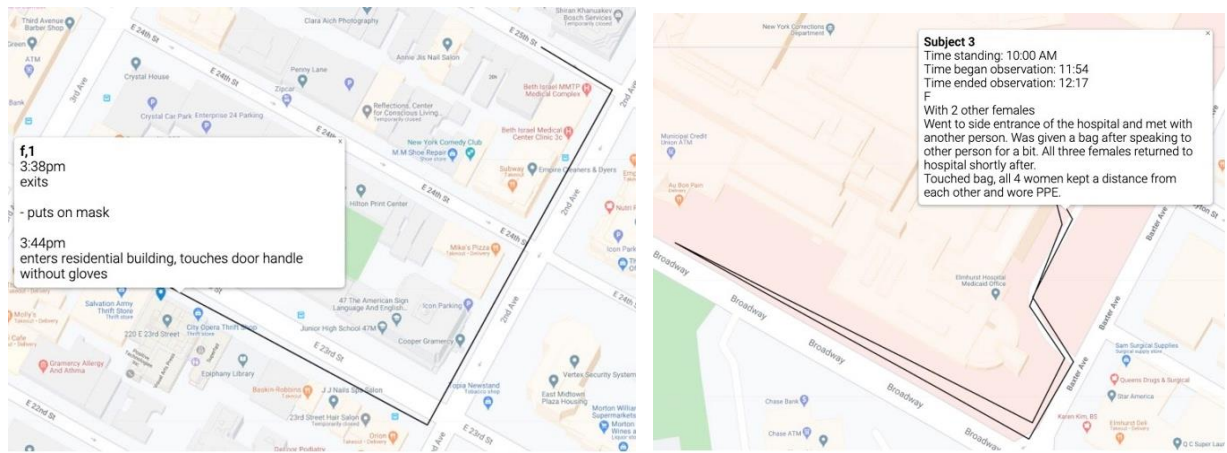

Figure 4. Sample records

The basic workflow of this dataset included five major steps. The first was exporting the smartphone records to either a Keyhole Markup Language (KML) or Keyhole Markup Language Zipped (KMZ) file depending upon the phone app. These formats are compatible with many geographic information systems (GISs), as well as Google-developed 3D visualization software including Google earth. The second stage was a manual extraction of the descriptive notes from the KML/KMZ file to a spreadsheet. To introduce quality control measures, these were scraped by one researcher and checked on an entry-by-entry level by a second researcher.

The third step involved introducing a secondary coding to many of the data fields, which allowed for a more generalized accounting of the data. For example, taxis, Ubers, and Lyfts were combined in the secondary coding as "vehicle for hire". Related activities also included the introduction of binaries (e.g. was the subject wearing PPE?). To ensure quality control, these were coded by one researcher and then checked on an entry-by-entry basis by a second researcher.

Step 4 was the finding and joining of zip code level weather and socio-economic data with each record. Finally, individual KML/KMZ files were transformed into a single shapefile. This was done first by generating multiple shapefiles with a Bash script transforming each individual file. These shapefiles were then integrated into a single shapefile that was then linked to the master comma-separated values (CSV) file to enable direct spatial analysis of the data.

The accompanying codebook explains on a column-by-column basis the variables in the master CSV file. The column name, variable name, label, description, missing value indicators, and data storage type (e.g. binary, string, character, numeric) for each variable are provided. The codebook also includes the methodology used for overcoming missing information. For example, in the master CSV file, there are columns for three pairs of different variables representing the start and end times. The first pair are in a character format recorded by the observer. The second pair uses the same format but provides estimated times for any records missing that information. This estimated time was computed with the MICE package in R using multiple imputations with 5 iterations. The final missing time variable was calculated by the average of 5 different results to minimize the standard error. Multiple imputations, in short, is a method to perform a simple imputation on a select variable (in this case, the time variable). This then replaces the variable with the prediction from the regression model with each iteration. Multiple imputations account for the model's uncertainty and sampling uncertainty. All imputations in this resource were based on "Missing at Random Assumptions", which indicates that the missing times were not randomly assigned. The third time instantiation has the start and end times appear as timestamps.

The master CSV file includes a total of 112 variables. Of these, 14 are the raw data parsed from the KML/KMZ files. These include the location, date, and start and end times of the record, as well as the subject's gender, number of accompanying persons, PPE usage, the destination (interim and final), up to 3 touched items, observer location, and additional comments. From those, a further 23 variables were derived. These include day of week, time of day, affiliated location information (e.g. address, borough), categorization of destinations and objects touched, facility type (hospital or urgent care), binaries on touching, PPE usage, returning to the facility, and use of mechanized transport. Based on the zipcode, date, and time, 7 local weather attributes obtained from the closest weather station were entered as variables. These include temperature, weather, wind speed, wind direction, humidity (possibility of raining), atmospheric pressure, and visibility. Based only on zip code, 62 variables with socio-economic data were included. The majority were taken from the American Community Survey (ACS) (2014-2019) with estimates of and margins of error provided in numbers and percentages for the total population, housing units, households, persons below poverty level, unemployment rate, per capita income, persons 25+ years without a high school diploma, persons 65 years or older, persons aged 17 and younger, noninstitutionalized population with disability, single parent household with children under 18, racial minorities (all persons non-white, non-Hispanic), and people age 5+ who speak English "not well" or "not at all"[4]. These are used by the Center for Disease Control (CDC) to calculate the Social Vulnerability index (SVI). Three variables related to socio-economic status were taken from United States Zip Codes[5] (https://www.unitedstateszipcodes.org/). These include population density, socioeconomic status, and average time to work. The remaining six variables are metadata.

**Data Resource Use**

Vector control strategies are central to the mitigation and containment of COVID-19, and have come in the form of municipal ordinances, which restrict the operational status of public and private spaces and associated municipal services. Having access to the documentation of real-world interactions between individuals, their immediate surroundings, and their destination choices upon exiting NYC healthcare facilities at the height of the city's spring COVID-19 outbreak can provide unprecedented insight into mechanisms from community and object-based transmission processes. This access is additionally essential for improving COVID-19 mitigation policy [6,7]. By establishing an empirical snapshot of the COVID-19 related vector environment surrounding these NYC healthcare facilities, highly detailed analysis is possible as per individual actions by gender and time of week in a range of communities variously impacted by COVID-19 at the onset of the first PAUSE order and subsequent mask mandate. Weather and socio-economic characteristics were joined to the data to provide location-based context in an effort to facilitate a wide-spread adoption of this data set. This rich data set will better position those in public health in their creation of effective interventions against community transmission going forward.

Because of the hyper-local, concurrent, multi-site collection efforts, there are many ways this data set could be used. One example would be to compare behaviors near facilities that are considered high risk (e.g. the hospitals many of which had temporary morgue trucks near the monitored exits) and facilities that are considered low risk (e.g. urgent care clinics). Such a study could consider whether there were statistically significant differences in behaviors at such locations or as a reflection of local community-based infection rates, hospitalization rates, or death rates. The data set can also be used to enhance models that are developed exclusively with footfall data extracted from cell phone signals (e.g. SAFEGRAPH). To date, no papers have been published by the authors on this data set. Work is presently being undertaken to look at how PPE usage, touch behavior, and destination selection changed over the nine-week observation period.

**Strengths**

The data set provides the basis for investigating the dynamics of disease distribution (e.g. diffusion theory) to explain transmission paths and mechanisms[8]. The data set presented herein contributes to this approach in documenting hyper-local, perishable data on healthcare egress patterns. Critically, as living in a pandemic is experientially new for most people, the hyper-local, perishable data contributes a unique and timely resource for the research and public health communities. As the data were collected over multiple weeks at multiple locations with distinct socio-economic characteristics and infection levels, more generalizable trends can be explored as well as the impact of both short-term and long-term conditions.

Specifically, when combined with complementary efforts to document 3D vector environments, the dataset provides an empirical basis for critical new disease transmission models and improvements to public health decision making, intervention, and risk communication. The data could also be included in probabilistic modeling and infrastructure disinfection planning for ongoing virus resurgence and mitigation efforts, as was predicted both by the US military and the World Health Organization[9,10] and are now fully in evidence in the US and Europe[11].

**Weaknesses**

While there are more than 5,100 records collected over the relatively short period, the number of records collected per day and per facility differ. Further, the observation period was not fully identical for every facility (e.g. not all began on March 22, 2020 and not all continued until May 19, 2020). Additionally, while most of the data collection process was highly standardized across the 16 observers, PPE usage was not collected at every facility. Finally, as records were not collected prior to PAUSE order and were only nine weeks in duration, there are neither robust pre-COVID behaviors nor long-term records against which to compare this data set.

**Data Resource Access**

All available data and the affiliated codebook are accessible through the NYU Faculty Digital Archive[12]. The codebook provides the variable lists, formats, and naming conventions. The KML/KMZ files can be opened in a wide range of free online viewers as well as any GIS compatible systems. The joint shape file must be opened with QGIS, ArcGIS, or a similar software. The data are freely available under the academic license posted with the data. All inquiries should be directed to the corresponding author.

**Acknowledgements**

This work was supported by the National Science Foundation (award #2027293), The Data Science and Software Services (DS3) which is funded by the Moore and Sloane foundations through the NYU Moore Sloane Data Science Environment, and Worth Sparks of Bluefield GIS.